\documentstyle[prd,aps,floats,tighten,epsfig,graphicx]{revtex}

\newcommand{\beq}{\begin{equation}}
\newcommand{\eeq}{\end{equation}}
\newcommand{\un}[1]{_{\scriptsize \mbox{#1}}} 

\newcommand{\etal}{{\rm et al.}}
\newcommand{\ie}{{\it i.e.}}

\newcommand{\msol}{{\rm M}_\odot}
\newcommand{\lsol}{{\rm L}_\odot}
\def\erf{\mathop{\rm erf}}

\newcommand{\app}[3]{Astropart.\ Phys.\ {\bf #1}, #3 (#2)} 

\newcommand{\hepph}[1]{{\tt hep-ph/#1}}
 
\newcommand{\prep}[3]{Phys.\ Rep.\ {\bf #1}, #3 (#2)} 
\newcommand{\plb}[3]{Phys.\ Lett.\ B\ {\bf #1}, #3 (#2)} 
\newcommand{\npb}[3]{Nucl.\ Phys.\ B\ {\bf #1}, #3 (#2)} 
\newcommand{\cpc}[3]{Comm.\ Phys.\ Comm.\ {\bf #1}, #3 (#2)} 
\renewcommand{\apj}[3]{Astrophys.\ J.\ {\bf #1}, #3 (#2)} 
\newcommand{\aj}[3]{Astronomical\ J.\ {\bf #1}, #3 (#2)} 
 
\renewcommand{\prl}[3]{Phys.\ Rev.\ Lett. {\bf #1}, #3 (#2)} 
\renewcommand{\prd}[3]{Phys.\ Rev.\ D\ {\bf #1}, #3 (#2)} 
\renewcommand{\rmp}[3]{Rev.\ Mod.\ Phys.\ {\bf #1}, #3 (#2)}

\newcommand{\href}[2]{#1}

\begin{document}
\draft


\title{Detection of neutralino annihilation photons from external 
galaxies} 

\author{E.~A.~Baltz}
\address{Department of Physics, University of California, Berkeley, 
CA 94720, USA}
\author{C.~Briot, P.~Salati, and R.~Taillet} \address{Laboratoire de 
Physique Th\'eorique LAPTH, BP110, F-74941 Annecy-le-Vieux Cedex, 
France}
\author{J.~Silk}
\address{Department of Astrophysics, Oxford University, Oxford OX1 
3RH, United Kingdom}
\maketitle

\smallskip
\vskip 0.5cm
\centerline{Draft Version of \today}

\begin{abstract}
We consider neutralino annihilation in dense extragalactic systems 
known to be dominated by dark matter, in particular M87 and several 
local dwarf spheroidal galaxies. These annihilations can produce 
energetic gamma rays which may be visible to atmospheric \v{C}erenkov 
telescopes. We explore the supersymmetric parameter space, and 
compute the expected flux of gamma--rays coming from these objects. 
It is shown that some parts of the parameter space lead to a signal 
observable with the next generation of \v{C}erenkov telescopes, 
provided the supersymmetric dark matter has a clumpy structure, as 
may be expected in a hierarchical scenario for structure formation.
\end{abstract}

\pacs{95.35.+d, 14.80.Ly, 95.85.Pw, 98.70.Rz} 

\section{Introduction}
\label{sec:introduction}

One of the favorite candidates for the astronomical missing mass is a 
neutral weakly interacting particle. Such a species is predicted in 
particular by supersymmetry, a theory that is actively tested at 
accelerators. It is conceivable therefore that most of the dark 
matter in the halo of the Milky Way is made of such particles. The 
mutual annihilation of these so--called neutralinos would yield, 
among a few other indirect signatures, a flux of high--energy 
gamma--rays. The latter has been extensively studied in the 
literature \cite{bengtsson90,berezinsky94,jkg96,bergstrom98}. It is 
unfortunately spoiled by the diffuse background produced by the 
spallation of cosmic--ray protons on the interstellar gas as 
discussed by Chardonnet \etal \cite{chardonnet95}. The distribution 
of molecular hydrogen inside the galactic ridge is not sufficiently 
well known to ensure an accurate prediction of the diffuse emission. 
This may obliterate a reliable interpretation of any putative 
gamma--ray excess in terms of supersymmetric dark matter. 

\vskip 0.2cm
We examine here the possibility of observing the gamma--ray signal 
from extra--galactic systems that contain large amounts of unseen 
matter. The giant elliptical galaxy M87, at the center of the Virgo 
Cluster, provides an excellent illustration. It is known to contain 
ten times more mass than our own Milky Way. Actually, X-ray 
observations of its ambient hot gas indicate that in the innermost 
100 kpc, the mass reaches $\sim 10^{13} \, \msol$. If neutralinos 
pervade this galaxy, a strong gamma--ray emission should be produced 
in the central region. Neutralino annihilation is associated to the 
gamma--ray flux at the Earth
\beq \Phi_{\gamma}^{\rm susy} \; = \; 
\frac{1}{4 \pi} \, \langle \sigma v \rangle \, N_{\gamma} \, 
{\displaystyle \int}_{\rm los} n_{\chi}^{2} \, ds
\label{gr_flux}
\eeq
which, in the case of M87, is two orders of magnitude larger than for 
own galaxy. On the other hand, M87 is quite distant, about 15 Mpc 
away. It should therefore appear as a bright gamma--ray spot on the 
sky, extending for at most a few hundred square arcminutes. The 
detection of such a strong but localized source is well--suited for 
atmospheric \v{C}erenkov telescopes (ACT) which can only monitor 
small portions of the sky at the same time but have very large 
effective collecting areas.

Dwarf spheroidal (dSph) galaxies are also suspected to contain 
significant amounts of dark matter. They orbit round the Milky Way 
and are intermediate in size between the globular clusters and larger 
systems such as the Magellanic clouds. The typical mass of dSph's is 
$\sim 1-4 \times 10^{7} \, \msol$ while their radii reach up to $\sim 
1-2$ kpc. They are also closer than M87. 

\vskip 0.2cm
The gamma--ray flux (\ref{gr_flux}) produced by annihilating species 
in M87 may be split into an astrophysical piece and a particle 
physics piece. The former term merely amounts to the integral along 
the line of sight of the dark matter density squared. It is 
approximately given by $M^{2} \, R^{-5}$ where $M$ denotes the mass 
contained within the radius $R$. In the case of M87, that line of 
sight integral is $\sim 10 \, {\msol}^{2}$ pc$^{-5}$ whereas it is 
$\sim 0.01 - 1 \, {\msol}^{2}$ pc$^{-5}$ for dwarf spheroidals. This 
compares to the Milky Way value of $\sim 0.1 \, {\msol}^{2}$ 
pc$^{-5}$ which corresponds to the inner 100 kpc. These rough 
estimates are improved in Sec.~\ref{sec:astrophysics} where the 
distribution of matter inside M87 and the dSph systems is modeled 
together with the radial profile of the neutralino induced gamma--ray 
emission. The various backgrounds to the latter are discussed in 
Sec.~\ref{sec:backgrounds} where a fiducial example is presented. In 
particular, the signal--to--noise ratio is derived as a function of 
the angular distance to the centers of the sources under scrutiny. 
The supersymmetric model is presented in Sec.~\ref{sec:susymodel}. We 
then delineate in Sec.~\ref{sec:sensitivity} the domain of neutralino 
masses and gamma--ray production cross sections which the next 
generation of atmospheric \v{C}erenkov telescopes will explore. Both 
continuum and monochromatic channels are featured. We finally draw 
some conclusions in Sec.~\ref{sec:conclusion} where we pay special 
attention to a possible clumpy structure of the neutralino 
distribution.

\section{The M87 and dSph Systems}
\label{sec:astrophysics}

The observed X-ray emission produced by clusters of galaxies such as 
Virgo results from the thermal bremsstrahlung that takes place in the 
hot diffuse intracluster gas (see \cite{sarazin} for a complete 
review). The temperature profile $T(r)$ of the gas can be inferred 
from the spectrum of the X--ray radiation. The electron number 
density profile $n_e (r)$ can be obtained from the X--ray surface 
brightness \beq
\Sigma(r) \; = \; {\displaystyle \int}
n_{e}(\sqrt{r^{2}+s^{2}}) \, L_{X} \, ds \;\; , \eeq
where $L_{X}$ denotes the X--ray luminosity of the plasma. The 
properties of the hot gas in the vicinity of M87 have been determined 
by Tsai \cite{tsai93}. The X--ray data are well fitted by
\beq
n_{e}(r) \; = \; n_{o} \,
\frac{\displaystyle (r/a_{1})^{-\alpha_{1}}} {\displaystyle 1 + 
(r/a_{1})^{\alpha_{3}}} \;\;\;\; \mbox{and} \;\;\;\;
T(r) \; = \; T_{\infty} \, \left(
\frac{\displaystyle r}{\displaystyle a_{2} + r} \right)^{\alpha_{2}} 
\;\; ,
\eeq
with $n_{0} = 4.31 \times 10^{-2} \, \mbox{cm}^{-3}$, $a_{1} = 6.63$ 
kpc, $a_{2} = 4.58 \times 10^{5}$ kpc, $\alpha_{1} = 0.49$, 
$\alpha_{2} = 0.114$, $\alpha_{3} = 0.869$ and $T_{\infty} = 8.35 
\times 10^{7}$ K. Assuming that the intracluster gas is in 
hydrostatic equilibrium, the total mass profile is given by
\beq
M(r) \; = \; - \,
\frac{kT \, r}{G \mu m_{p}} \,
\left( \frac{d \log n_{e}}{d \log r} + \frac{d \log T}{d \log r} 
\right) \;\; ,
\label{eq:mtot}
\eeq
where $\mu m_{p}$ is the average mass of a particle in the gas. The 
total mass inside M87 reaches $8.4 \times 10^{11} \, \msol$ at 10 kpc 
and $1.4 \times 10^{13} \, \msol$ at 100 kpc. The contribution from 
the stars alone was obtained from B-band surface brightness profiles 
(see \cite{tsai93} and references therein). The dark matter component 
distribution is derived once the stars and the gas have been removed. 
Mappings of the X--ray emission in the Virgo cluster are consistent 
with the hypothesis that this system and its central galaxy contain 
non--baryonic dark matter. The ratio of the dark matter density to 
the baryonic matter density is also borrowed from \cite{tsai93} \beq
{\displaystyle \frac{\rho_{\rm DM}}{\rho_{\rm B}}} \; = \; 0.15 \,
\left( \frac{r}{1 \, {\rm kpc}} \right)^{1.5} \;\; . \eeq
The integral along the line of sight of $\rho_{\rm DM}^{2}$ is 
readily obtained.
\vskip 0.2cm

\begin{table}[h!]
\[
\begin{array}{ l c c c c c c c c c}
\hline \hline
\\
& &
L &
M/L &
M &
d &
\sigma_{\rm obs} &
c &
r_{t} &
\sigma_{\rm th}
\\
{\rm Name} & &
(10^{5} \; \lsol) &
(\msol / \lsol) &
(10^{6} \; \msol) &
({\rm kpc}) &
({\rm km/s}) &
&
({\rm kpc}) &
({\rm km/s})
\\
\\
\hline
\\
{\rm Carina} & & 2.4 \pm 1.0 & 59 \pm 47 & 14.1 \pm 10.0 & 85 \pm 5 & 
6.8 \pm 1.6 & 0.5 & 1.67 & 7.6 \\
{\rm Draco} & & 1.8 \pm 0.8 & 245 \pm 155 & 44.1 \pm 24.0 & 72 \pm 3 
& 10.2 \pm 1.8 & 0.5 & 2.18 & 11.8 \\
{\rm Ursa \; Minor} & & 2.0 \pm 0.9 & 95 \pm 43 & 19.0 \pm 10.5 & 64 
\pm 5 & 12.0 \pm 2.4 & 0.5 & 0.80 & 12.8 \\
{\rm Sextans} & & 4.1 \pm 1.9 & 107 \pm 72 & 43.8 \pm 23.6 & 83 \pm 9 
& 6.2 \pm 0.8 & 1 & 3.02 & 8.8
\\
\\
\hline \hline
\end{array} \]
\caption{These four dSph galaxies have very large M/L--values and are within 
100 kpc from the Milky Way center. The luminosities $L$, 
mass--to--light ratios, total masses $M$, galactocentric distances 
$d$ and concentrations $c$ have been borrowed from [10].
The velocity dispersions $\sigma_{\rm obs}$ are those given by 
[13]
. These systems have been modeled with King profiles as 
explained in the text. Consistency may be checked by comparing the 
resulting velocity dispersions $\sigma_{\rm th}$ with the observed 
values $\sigma_{\rm obs}$. Also indicated are the derived tidal radii 
$r_{t}$, in kpc.}
\label{table:dSph}
\end{table}

Dwarf spheroidal (dSph) galaxies also seem to contain large amounts 
of unseen material \cite{burkert97}. Like globular clusters, dSph's 
have low luminosities of order $10^{5} - 10^{7} \, \lsol$ within an 
ill-defined radius $r_{t}$ of a few kpc. Their structure also follows 
a King profile (see \cite{gallagher94} for a review). If these 
systems are in virial equilibrium, a velocity dispersion $\sigma$ of 
a few km/s translates into a mass \beq
M_{\rm dSph} \simeq 2.3 \times 10^{7} \, \msol \; \left( 
\frac{\sigma}{10 \, {\rm km/s}} \right)^{2} \; \left( \frac{1 \, {\rm 
kpc}}{r_{t}} \right) \;\; , \eeq
hence these systems have mass--to--light ratios that may reach $\sim$ 
200. We have been interested here in four galaxies with exceptionally 
high M/L--values \cite{irwin95} and galactocentric distances less 
than 100 kpc. The dSph's featured in table~\ref{table:dSph} are 
Carina, Draco, Ursa Minor and Sextans. %
We have modeled their mass distributions with a one--component King 
profile \cite{king65} for which the phase space density is given by 
\beq
f(E) \, \propto \,
\exp \left( \frac{\phi_{t} - E}{\sigma^{2}} \right) \; - \; 1 \;\; .
\label{phase_space_density}
\eeq
The stellar or particle energy per unit mass is denoted by $E = \phi 
+ v^{2}/2$ while $\phi_{t} = \phi(r_{t})$ is the potential at the 
tidal radius $r_{t}$ of the dSph. Beyond that boundary, stars have 
enough energy to escape from the gravitational field of the system 
and are captured by the tidal field of the nearby Milky Way. The mass 
distribution ensues from the phase space density 
(\ref{phase_space_density})
\beq
\rho (r) \, \propto \,
\frac{\sqrt{\pi}}{4} \exp({\displaystyle u}) \erf(\sqrt{u}) \, - \, 
\frac{\sqrt{u}}{2} \, - \, \frac{u^{3/2}}{3} \;\; ,
\eeq
where $u$ denotes the ratio $\left\{ \phi_{t} - \phi(r) \right\} / 
\sigma^{2}$ while erf is the error function. One component King 
models are completely determined once the velocity dispersion 
$\sigma$, the central density $\rho_{c}$ and the ratio $\chi = 
\left\{ \phi_{t} - \phi (0) \right\} / \sigma^{2}$ are specified. Any 
combination of these quantities suffices. %
The concentrations of the four dSph's under scrutiny are given in 
table~\ref{table:dSph} as determined by \cite{irwin95}. They are 
related to the tidal and core radii through
\beq
c \; = \; \log_{10} \left( \frac{r_{t}}{r_{c}} \right) \;\; , \eeq
where the core radius $r_{c}$ is defined as \beq
r_{c} \; = \;
{\displaystyle \frac{3 \sigma}{\sqrt{4\pi G \rho_{c}}}} \;\; . \eeq
Values of the concentration of 0.5 and 1 respectively translate into 
the ratios $\chi = 1.97$ and 4.85. Fixing the concentration allows 
for the determination of the density profile $\rho(r) / \rho_{c}$ as 
a function of the reduced radius $r / r_{c}$. The second input to our 
calculations is the dSph mass as given by the central values of 
table~\ref{table:dSph}. We finally set the average dSph density 
$\bar{\rho}_{\rm dSph}$ by requiring that the proper gravitational 
field of these galaxies is compensated, at their boundaries, by the 
tidal field of the Milky Way taken at perigalacticon. Assuming a 
logarithmic galactic potential implies that
\beq
\bar{\rho}_{\rm dSph} \; = \; f(e) \, \bar{\rho}_{\rm G} \;\; ,
\label{tidal_relation}
\eeq
where the function $f(e)$ depends on the orbital eccentricity of the 
satellite galaxy \cite{king62}
\beq
f(e) \; = \;
{\displaystyle \frac{1 \, + \,
\left[ \left(1 + e \right)^{2} / 2e \right] \, \ln \left[ \left( 1+e 
\right) / \left( 1-e \right) \right]} {\left( 1 - e \right)^{2}}} 
\;\; .
\eeq
The average galactic density $\bar{\rho}_{\rm G}$ is understood 
within the sphere centered on the Milky Way, with radius equal to the 
semimajor axis $a$ of the dSph orbit. The corresponding mass is given 
by \beq
M_{\rm G} \; = \; 1.1 \times 10^{10} \msol \times a \, \left[ 
\mbox{kpc} \right] \;\; .
\eeq
We set the semimajor axis $a$ equal to the dSph galactocentric 
distance $d$. Eccentricities in the range between 0 and 1/2 are 
generally assumed. Here, their values were derived by requiring that 
the resulting velocity dispersions $\sigma_{\rm th}$ should be close 
to the observations -- see table~\ref{table:dSph}. In the case of 
Carina and Draco, the eccentricity is $e = 0$ while for Ursa Minor, a 
value of $e = 1/2$ is appropriate. As regards Sextans, we simply 
imposed the equality between the dSph and galactic average densities, 
\ie, $\bar{\rho}_{\rm dSph} \, = \, \bar{\rho}_{\rm G}$, in order to 
get a velocity dispersion of 8.8 km/s, not too far from the measured 
value of $6.2 \pm 0.8$ km/s. Assuming relation~(\ref{tidal_relation}) 
and a circular orbit would have lead to $\sigma_{\rm th} \, = \, 9.9$ 
km/s.
%
In modeling the dSph galaxies, we have adopted the point of view that 
these satellites contain large amounts of dark matter as implied by 
their high velocity dispersions. In the alternative explanation, the 
latter are the result of an ongoing tidal stripping. However, Oh 
\etal \cite{Oh95} showed that unbound but not yet dispersed systems 
have velocity dispersions fairly close to the values derived from the 
virial equilibrium. The large M/L--values of the four dSph systems at 
stake do not depend therefore on whether these galaxies are in 
equilibrium or are currently being torn apart.
Notice finally that most of the putative neutralino--induced 
gamma--ray signal emitted by these dSph's is produced in the inner 10 
arcmin as shown in the right panel of Fig.~\ref{fig:profile}. It 
weakly depends on the actual boundaries of these systems.

\section{The Gamma-ray Signal and its Backgrounds} 
\label{sec:backgrounds}

The gamma--ray flux produced by a halo of dark matter particles may 
be computed as follows. First, for each set of supersymmetric 
parameters, we derived the thermally averaged annihilation cross 
section $\langle \sigma v\rangle_{A}$ into some channel $A$. Then, we 
used the Lund Monte Carlo \cite{pythia} to compute the photon 
spectrum per annihilation from hadronization or decay processes, and 
we integrated over some specified threshold which depends on the 
atmospheric \v{C}erenkov telescopes in question. Finally, we have 
summed over all channels to find the total photon spectrum from a 
given supersymmetric model.
The volume production rate of gamma--rays is simply given by 
$n_{\chi}^{2} \, \langle \sigma v \rangle \, N_{\gamma}$, where 
$n_{\chi}$ is the number density of annihilating particles and 
$N_{\gamma}$ is the mean number of photons above the threshold. The 
corresponding flux at the Earth depends on the integral, along the 
line of sight, of the dark matter density squared
\beq
\Phi_{\gamma}^{\rm susy} \; = \;
\frac{1}{4 \pi} \,
{\displaystyle
\frac{\langle \sigma v \rangle \, N_{\gamma}}{m_{\chi}^{2}} } \,
{\displaystyle \int}_{\rm los} \rho_{\chi}^{2} \, ds \;\; , \eeq
as well as on the gamma--ray production cross section $\langle \sigma 
v \rangle \, N_{\gamma}$ and the mass $m_{\chi}$ of the species. 

\vskip 0.2cm
Atmospheric \v{C}erenkov telescopes (ACT) offer the most promising 
method to detect gamma--rays with energies above 100 GeV. When 
passing through the atmosphere, incoming high--energy photons emit a 
\v{C}erenkov radiation that is detected and analyzed to infer the 
energy and arrival direction of the primary gamma--ray. The 
performances of ACTs are determined by the surface of the optical 
reflectors used to collect the \v{C}erenkov light, and the ability to 
reject the cosmic--ray induced background. A thorough analysis of 
these performances can be found in Aharonian \cite{aharonian}. We 
will simply model them by the effective collecting area $A_{\rm 
eff}$, a feature of the instrument that takes into account in 
particular the actual surface of detection together with the energy 
dependent detection efficiency. 

\begin{figure*}[!h]
\centerline{\epsfig{file=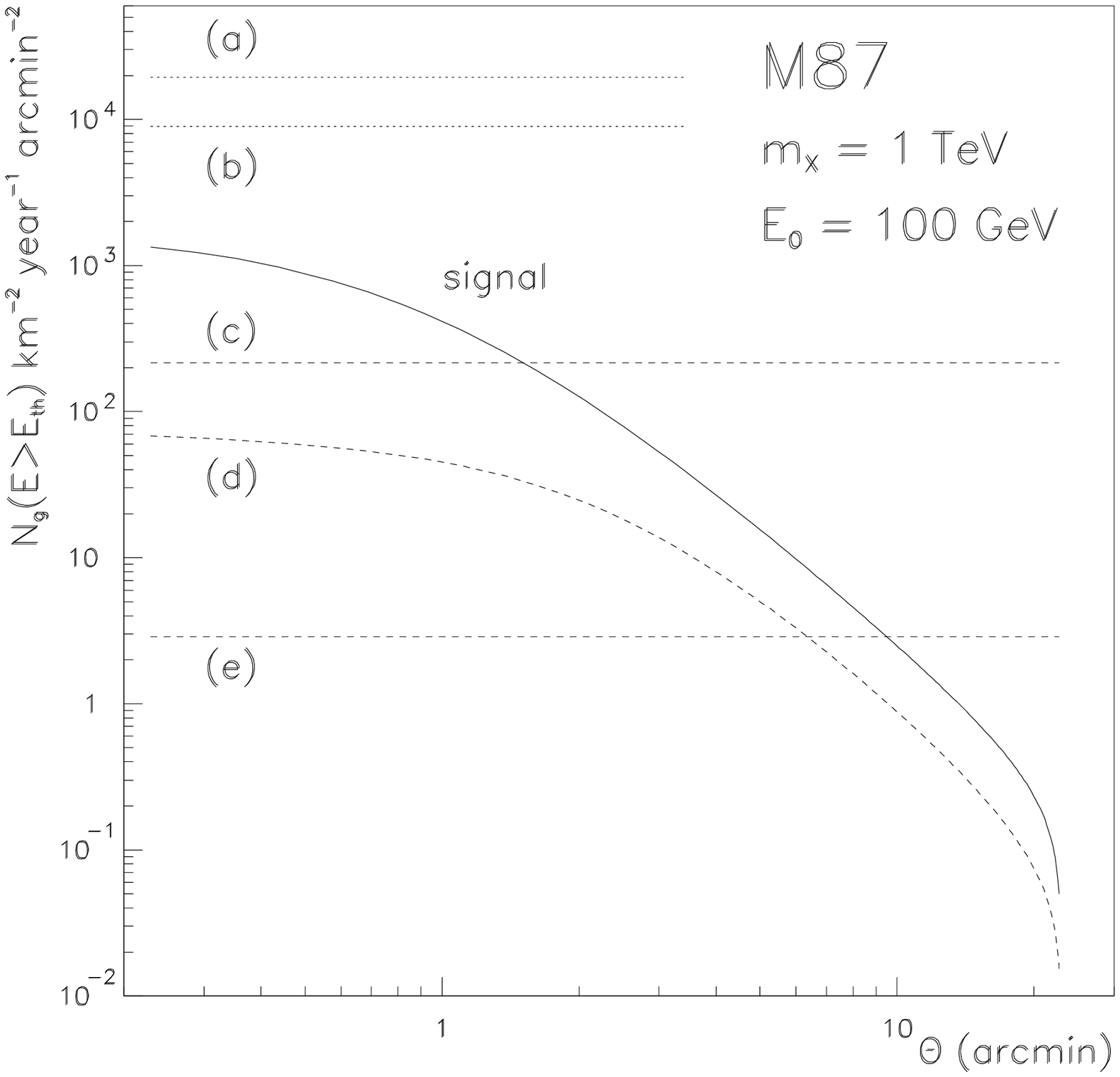,width=0.5\textwidth} 
\epsfig{file=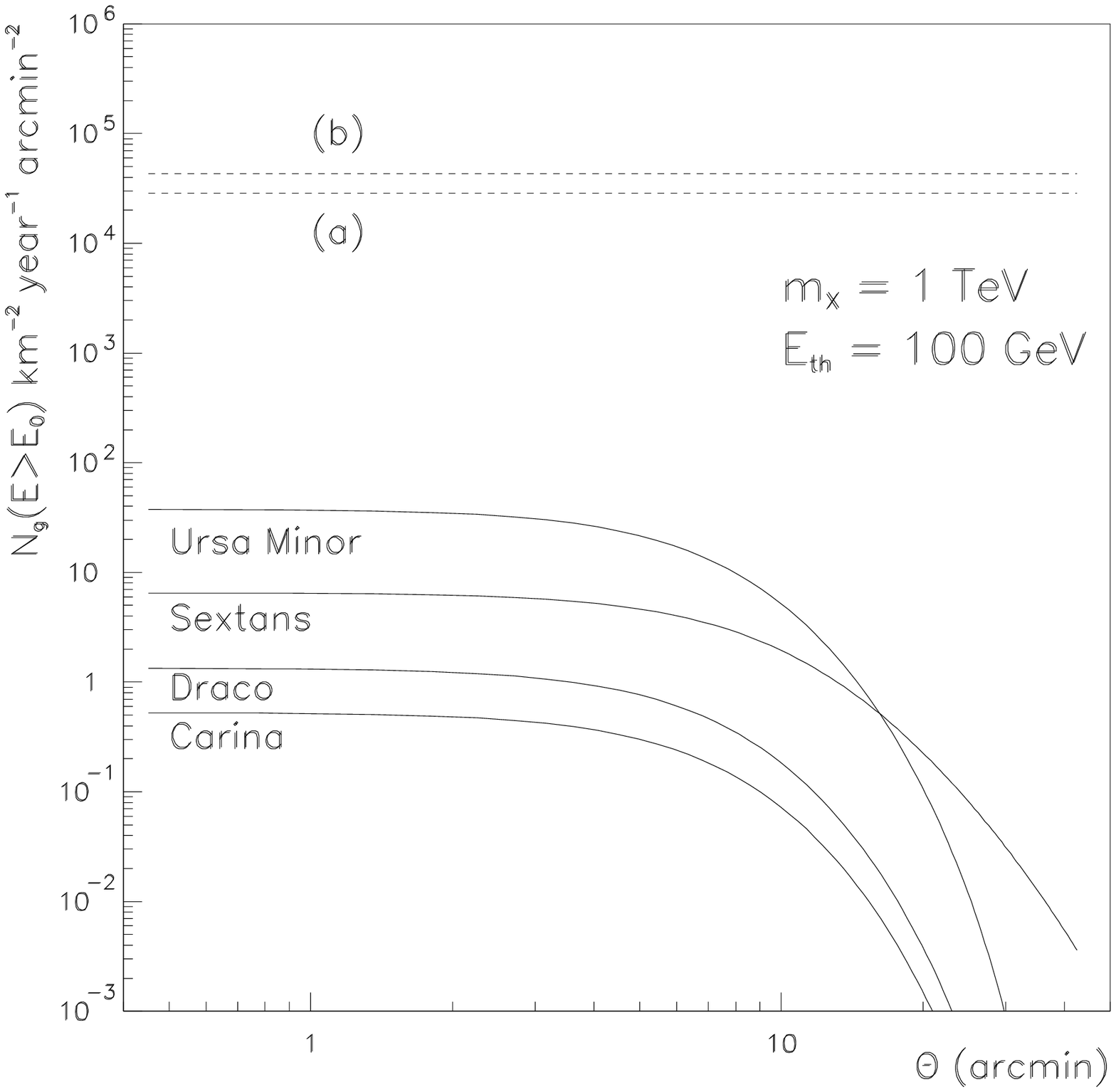,width=0.5\textwidth}} \caption{
The radial profiles of the neutralino--induced signal (solid curves) 
and of the various backgrounds (dotted and dashed lines) are plotted 
as a function of the angular distance to the source centers. A 
fiducial model with $m_{\chi}$ = 1 TeV and
$\langle \sigma v \rangle_{\rm cont.} N_\gamma = 10^{-25}$ cm$^{3}$ 
s$^{-1}$ is taken while a threshold of 100 GeV is assumed. In the 
left panel, the backgrounds are respectively labeled as
({\em a}): electronic;
({\em b}): hadronic;
({\em c}): extragalactic;
({\em d}): M87 and
({\em e}): Milky Way gamma--ray diffuse emissions. The right panel 
shows nearby dwarf spheroidal galaxies with exceptionally high 
M/L--values. The neutralino--induced signal is flat within 10 arcmin 
from the center. The total background is featured by the dashed line 
(g). Carina is the exception with a slightly larger value (f) 
dominated by the Milky Way diffuse emission. That system lies within 
the galactic ridge. }
\label{fig:profile}
\end{figure*}

\vskip 0.2cm
The surface brightness $\mu_{\gamma}^{\rm susy}$ of the source as 
seen by an ACT depends on the collecting area of the instrument and 
on the integration time $T$
\beq
\mu_{\gamma}^{\rm susy} \; = \;
\Phi_{\gamma}^{\rm susy} \, A_{\rm eff} \, T \;\; . \eeq
We plot the surface brightness in Fig.~\ref{fig:profile} for a 
collecting area $A_{\rm eff}$ = 1 km$^{2}$ and for an entire year of 
effective observation. In the fiducial model taken here, the 
neutralino mass is $m_{\chi} = 1$ TeV while the annihilation cross 
section has been set equal to $\left< \sigma v \right> \, N_{\gamma} 
\, = \, 10^{-25}$ cm$^{3}$ s$^{-1}$. The surface brightness 
corresponds to the number of photons above a threshold of 100 GeV 
that are produced per square arcmin of the source. In the left panel, 
the giant galaxy M87 is featured whereas in the right panel, the four 
dSph satellites discussed in the previous section, \ie, Carina, 
Draco, Sextans and Ursa Minor are presented. The signal declines away 
from the centers of these systems. It follows the integral along the 
line of sight of the neutralino density squared -- remember that 
$n_{\chi} = \rho_{\chi} / m_{\chi}$. Actually, we must take into 
account the finite angular resolution of the instrument. When the 
telescope is pointed towards the center of the object, photons coming 
from directions within the resolution cone also reach the detector. 
Thus, the relevant quantity to examine is the integral of the surface 
brightness $\mu_{\gamma}^{\rm susy}$ over a disk centered on the 
source, with angular radius $\theta$
\beq
N_{\rm s} \; = \;
{\displaystyle \int_{0}^{\theta}} \, 2 \pi \alpha \, 
\mu_{\gamma}^{\rm susy} \, d\alpha \;\; . \eeq
When the angular resolution of the ACT is good enough, the size of 
that circular region is generally specified by the requirement that 
the signal--to--noise ratio should be optimal. 

\vskip 0.2cm
The number of neutralino--induced source photons $N_{\rm s}$ should 
actually be compared to the background $N\un{bg}$, the origins of 
which are various. Would it be constant, this background could easily 
be removed from the signal. Actually, it follows Poisson statistics, 
and as such, it exhibits fluctuations
of amplitude $\sqrt{N\un{bg}}$. These fluctuations have a smaller 
chance of being interpreted as a signal when the significance \beq
S \; = \; \frac{N_{\rm s}}{\sqrt{N\un{bg}}} \eeq
is large. There are two main types of background events. %
First, an experimental background is due to hadronic and electronic 
cosmic--rays that impinge on the top of the atmosphere. The induced 
showers can be misinterpreted as gamma--ray events. Electrons make up 
the largest source of background insofar as their showers are of the 
electromagnetic type and cannot be disentangled from those generated 
by the impact of high--energy photons. The corresponding flux steeply 
decreases at high energy \cite{nishimura}
\beq
\Phi_{\rm e} \; = \; \left( 6.4 \times 10^{-2} \; {\rm GeV^{-1} \, 
cm^{-2} \, s^{-1} \, sr^{-1}} \right) \, \left( \frac{E}{\rm 1 \, 
GeV} \right)^{-3.3 \pm 0.2} \;\; , \eeq
and leads to the background
\beq
\frac{d N\un{e}}{d \Omega} (E>E_{0}) \; = \; \left( 1.9 \times 10^{4} 
\; {\rm km^{-2} \, yr^{-1} \, arcmin^{-2}} \right) \, \left( 
\frac{E_0}{\mbox{100 GeV}} \right)^{-2.3} \;\; . \eeq
In the left panel of Fig.~\ref{fig:profile}, that electron induced 
background is featured by the dotted line (a). It is more than an 
order of magnitude larger than the neutralino--induced signal at 
maximum. It is also noticeably flat all over the source.
Observations performed between 50 GeV and 2 TeV yield a hadron flux 
\cite{ryan}
\beq
\Phi_{\rm had} \; = \; \left( 1.8 \;
{\rm GeV^{-1} \, cm^{-2} \, s^{-1} \, sr^{-1}} \right) \, \left( 
\frac{E}{\rm 1 \, GeV} \right)^{-2.75} \;\; . \eeq
Hadron--induced showers are more extended on the ground than those of 
the electromagnetic type. Stereoscopy is a powerful tool to 
discriminate hadrons from electrons and gamma--rays. The CAT 
experiment, for instance, has already achieved a rejection factor of 
one misidentified event over a sample of 600 showers generated by 
cosmic--ray hadrons \cite{nuss}. Here, we have assumed an even better 
rejection factor, with only one misidentified hadron out of a 
thousand showers. This yields a hadron background \beq
\frac{d N\un{had}}{d \Omega} (E>E_{0}) \; = \; \left( 8.7 \times 
10^{3} \; {\rm km^{-2} \, yr^{-1} \, arcmin^{-2}} \right) \, \left( 
\frac{E_0}{\mbox{100 GeV}} \right)^{-1.75} \;\; . \eeq
which corresponds to the dotted line (b) of Fig.~\ref{fig:profile}. 
Once again, that background is flat over M87. %
Next come the astrophysical sources of background. To commence, 
Sreekumar \etal \cite{sreekumar} have measured an extragalactic 
component in the gamma--ray diffuse emission with the EGRET 
instrument on board the Compton gamma--ray observatory \beq
\Phi_{\rm eg} \; = \; \left( 7.32 \pm 0.34 \times 10^{-9} \; {\rm 
MeV^{-1} \, cm^{-2} \, s^{-1} \, sr^{-1}} \right) \, \left( 
\frac{E}{\rm 451 \, MeV} \right)^{-2.10 \pm 0.03} \;\; . \eeq
This translates into the background (c)
\beq
\frac{d N\un{eg}}{d \Omega} (E>E_{0}) \; = \; \left( 2.1 \times 
10^{2} \; {\rm km^{-2} \, yr^{-1} \, arcmin^{-2}} \right) \, \left( 
\frac{E_0}{\mbox{100 GeV}} \right)^{-1.1} \;\; . \eeq
In our fiducial example, the neutralino--induced signal exceeds the 
extragalactic background within $\sim$ 1 arcmin from the center of 
M87.
Then, we have modeled the gamma--ray diffuse emission from that giant 
elliptical galaxy itself. Local cosmic--rays interact with the 
cluster gas to produce high--energy photons that may potentially 
contaminate the signal. In the inner 10 kpc, the magnetic field of 
M87 is comparable to that of the Milky Way, with a magnitude $\sim$ 1 
$\mu$G. It falls by a factor of ten outwards in the Virgo cluster, at 
a distance of $\sim$ 100 kpc. Assuming that equipartition of energy 
holds between this magnetic field and the cosmic--rays that pervade 
M87 -- just like in our own galaxy -- we infer a gamma--ray 
emissivity of
\beq
I_{\rm H}(E) \; = \; \left( 2 \times 10^{-35} \; {\rm GeV^{-1} \, 
s^{-1} \, sr^{-1}} \right) \, \left( \frac{E}{\rm 1 \, TeV} 
\right)^{-2.73} \;\; , \label{emissivity_inner_M87}
\eeq
per hydrogen atom illuminated by high--energy protons. Once 
multiplied by the hydrogen column density across M87, it yields the 
in situ diffuse gamma--ray flux. We have assumed that hydrogen is the 
main constituent of the gas at the center of the Virgo cluster so 
that we have integrated the electron
density, as derived in Sec.~\ref{sec:astrophysics}, along the 
appropriate lines of sight. The gamma--ray emissivity as given by 
relation (\ref{emissivity_inner_M87}) is typical of the inner 10 kpc 
inside M87. It has been rescaled by the factor
$\left\{ 1 \, + \, \left( r / 10 \, {\rm kpc} \right)^{2} 
\right\}^{-1}$ to account for the outward decrease of both the 
magnetic energy and cosmic--ray flux. The M87 gamma--ray diffuse 
background is presented as the dashed line (d) of 
Fig.~\ref{fig:profile}. It falls outwards as a result of the combined 
decrease
of the hydrogen column density and of the cosmic--ray flux. %
Finally, we have taken into account the Milky Way diffuse emission. 
We use hydrogen column densities inferred from the dust maps of 
\cite{dustmap}. A hydrogen column density of $1.69 \times 10^{20}$ H 
cm$^{-2}$ in the direction of M87 corresponds to
\beq
\frac{d N\un{MW}}{d \Omega} (E>E_{0}) \; = \; \left( 2.8 \; {\rm 
km^{-2} \, yr^{-1} \, arcmin^{-2}} \right) \, \left( 
\frac{E_0}{\mbox{100 GeV}} \right)^{-1.73} \;\; , \eeq
and to the dashed line (e). The Milky Way component is the weakest 
source of background. Electrons and in a lesser extent hadrons -- as 
long as rejection is efficient -- dominate.
In the right panel of Fig.~\ref{fig:profile}, the total background is 
presented. It encompasses the various sources discussed above except 
a local diffuse emission. The dSph galaxies contain old stars and the 
acceleration mechanisms at work in the Milky Way are presumably less 
prevalent there. Even in the case of the brightest source, Ursa 
Minor, the signal is still three orders of magnitude below the 
background -- see curve (g). Because Carina lies in the sky towards 
the galactic ridge, the hydrogen column density of the Milky Way is 
quite large, reaching a value of $8.7 \times 10^{23}$ H cm$^{-2}$. 
The gamma--ray diffuse emission of our own galaxy dominates the 
background (f) in that specific case.

\begin{figure*}[!h]
\centerline{\epsfig{file=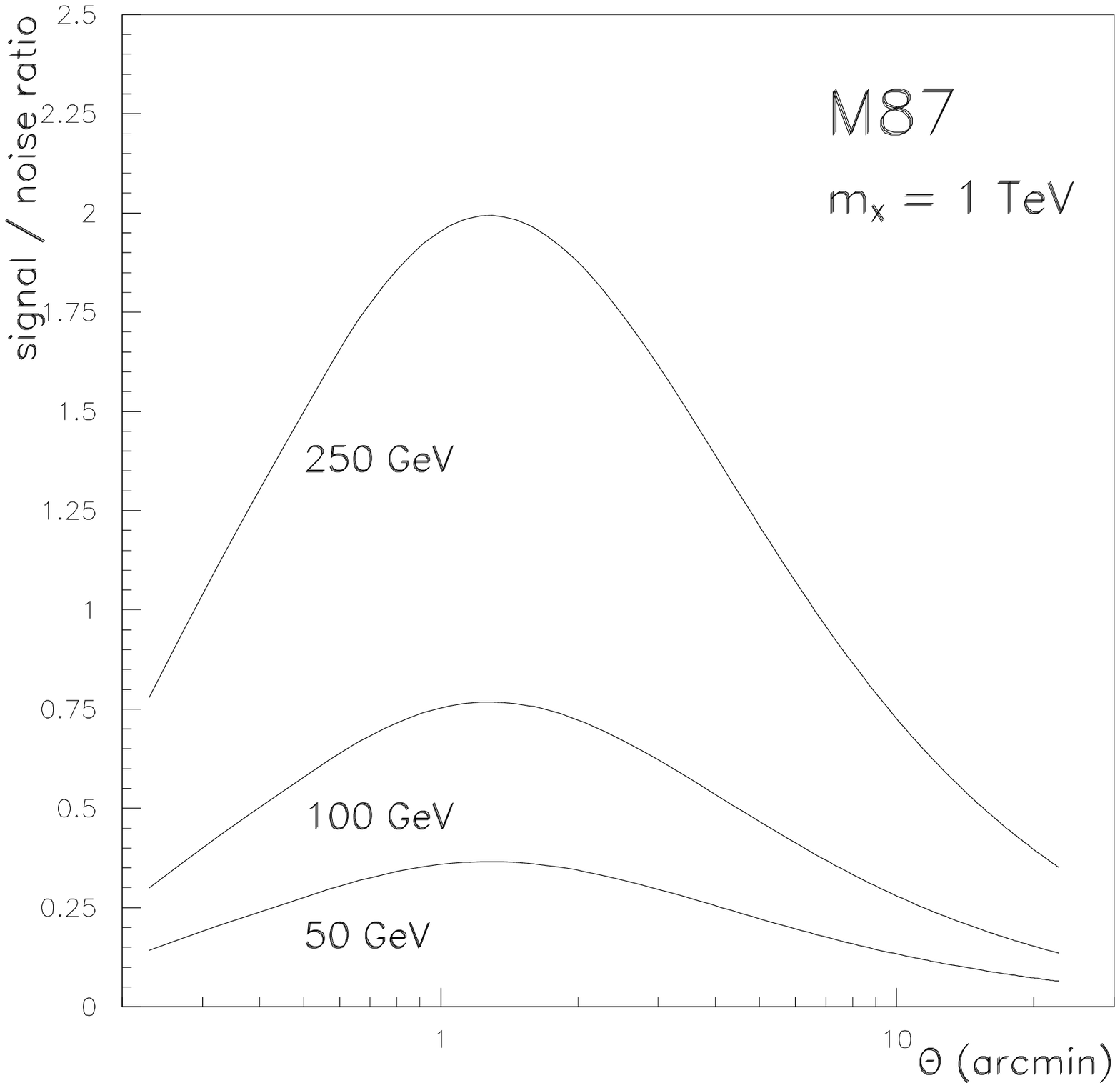,width=0.5\textwidth} 
\epsfig{file=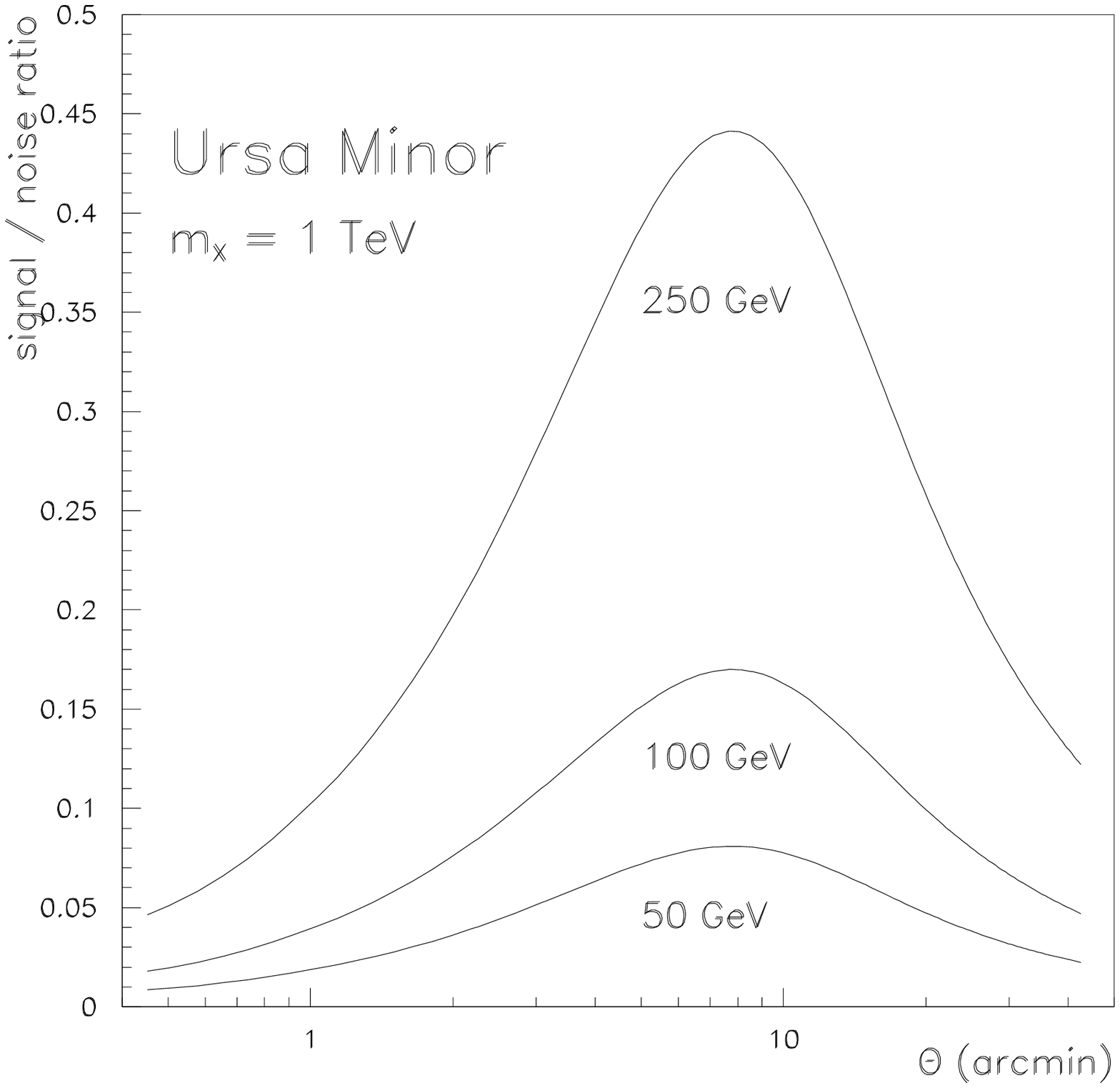,width=0.5\textwidth}} \caption{
Signal--to--noise ratio as a function of threshold and beam size. 
Again a fiducial model with $m_{\chi}$ = 1 TeV and $\langle \sigma v 
\rangle_{\rm cont.} \, N_{\gamma} \, = \, 10^{-25}$ cm$^{3}$ s$^{-1}$ 
is taken, while a more realistic integration of 0.01 km$^2$ yr
is now assumed. The left panel features the case of M87 for which the 
significance $S$ peaks at an angular radius of $\sim$ 1.4 arcmin. For 
the brightest dSph Ursa Minor (right panel), the optimal beam size is 
7.7 arcmin. Three values of the threshold energy are presented. }
\label{fig:signal_to_noise}
\end{figure*}

\vskip 0.2cm
The significance $S$ is presented in Fig.~\ref{fig:signal_to_noise} 
for M87 (left panel) and in the case of Ursa Minor, the best dSph 
source (right panel). Should the signal be flat, both $N_{\rm s}$ and 
$N\un{bg}$ would be proportional to the surface of the sky monitored 
by the ACT. As a result, the signal--to--noise ratio would just 
increase like the angular radius $\theta$ of the observed region. In 
Fig.~\ref{fig:signal_to_noise}, it actually increases
as the beam size opens up. Because the signal is not flat but weakens 
far from the source, the signal--to--noise ratio reaches a maximum 
and drops at larger radii.
The ACT acceptance has been set equal to 0.01 km$^{2}$ yr. The 
neutralino mass is still 1 TeV whereas the gamma--ray production 
cross section $\langle \sigma v \rangle_{\rm cont.} \, N_\gamma \, = 
\, 10^{-25}$ cm$^{3}$ s$^{-1}$ is the same for the three values of 
the threshold energy featured
here. As the background is flat over the source, the 
signal--to--noise ratio always
peaks at the same angular position. For M87, the significance is the 
largest for an
angular aperture of $\sim$ 1.4 arcmin. Because Ursa Minor is closer, 
the optimal
beam size becomes 7.7 arcmin. As both the electronic and hadronic 
backgrounds weaken with the gamma--ray energy, the magnitude of the 
maximum increases with the threshold as is clear in both panels. For 
M87 and a 50 GeV threshold, the significance reaches a value of 
$\sim$ 0.4 whereas for 250 GeV, the peak value increases up to 2. In 
the case of Ursa Minor, the variations of the significance are just 
the same. Should the acceptance be 1 km$^{2}$ yr, these 
signal--to--noise ratios would be rescaled up by an order of 
magnitude.

\section{Exploring the SUSY Parameter Space} \label{sec:susymodel}

We have explored the Minimal Supersymmetric Standard Model (MSSM)\@. 
This framework has many free parameters, but with reasonable 
assumptions the set of parameters is reduced to seven: the Higgsino 
mass parameter $\mu$,
the gaugino mass parameter $M_{2}$,
the ratio of the Higgs vacuum expectation values $\tan \beta$, the 
mass of the $CP$--odd Higgs boson $m_{A}$, the scalar mass parameter 
$m_{0}$
and the trilinear soft SUSY--breaking parameters $A_{b}$ and $A_{t}$ 
for third generation squarks.
The only constraint from supergravity that we imposed is gaugino mass 
unification,
though the relaxation of this constraint would not significantly 
alter our results.
For a more detailed description of these models, see 
Refs.~\cite{coann,jephd}. 

\vskip 0.2cm
The lightest stable supersymmetric particle in most models is the 
lightest of the neutralinos, which are superpositions of the 
superpartners of the gauge and Higgs bosons. We used the one--loop 
corrections for the neutralino and chargino
masses given in Ref.~\cite{neuloop}. For the Higgs bosons we used the 
leading log two--loop radiative corrections, calculated within the 
effective potential approach given in Ref.~\cite{carena}. We made 
thorough random scans of the model parameter space, with overall 
limits of the seven MSSM parameters as given in Table~\ref{tab:scans}.

\begin{table}
\begin{tabular}{rrrrrrrr}
Parameter & $\mu$ & $M_{2}$ & $\tan \beta$ & $m_{A}$ & $m_{0}$ & 
$A_{b}/m_{0}$ & $A_{t}/m_{0}$ \\
Unit & GeV & GeV & 1 & GeV & GeV & 1 & 1 \\ \hline Min & -50000 & 
-50000 & 1.0 & 0	& 100 & -3 & -3 \\
Max & 50000 & 50000 & 60.0 & 10000 & 30000 & 3 & 3 \\ 
\end{tabular}
\caption{
The ranges of parameter values used in our scans of the MSSM 
parameters. In total, we used approximately 116,000 models that are 
not excluded by current accelerator constraints. }
\label{tab:scans}
\end{table}

\vskip 0.2cm
Each model was examined to see if it is excluded by the most recent 
accelerator constraints. The most important of these are the LEP 
bounds \cite{lepbounds} on the lightest chargino mass
\beq
m_{\chi_{1}^{+}} > \left\{
\begin{array}{lcl}
91 {\rm ~GeV} & \quad , \quad &
| m_{\chi_{1}^{+}} - m_{\chi^{0}_{1}} | > 4 {\rm ~GeV} \\ 85 {\rm 
~GeV} & \quad , \quad & {\rm otherwise} \end{array} \right.
\eeq
and on the lightest Higgs boson mass $m_{H_{2}^{0}}$ (which ranges 
from 72.2--88.0 GeV depending on $\sin (\beta - \alpha)$ with 
$\alpha$ being the Higgs mixing angle) and the constraints from $b 
\to s \gamma$ \cite{cleo}. 

\vskip 0.2cm
For each allowed model, the relic density of neutralinos 
$\Omega_{\chi} h^2$ was calculated, where $\Omega_{\chi}$ is the 
density in units of the critical density
and $h$ is the present Hubble constant in units of $100$ km s$^{-1}$ 
Mpc$^{-1}$\@.
We used the formalism of Ref.~\cite{GondoloGelmini} for resonant 
annihilations, threshold effects, and finite widths of unstable 
particles and we included all two--body
tree--level annihilation channels of neutralinos. We also included 
the so--called
coannihilation processes between all neutralinos and charginos in the 
relic density
calculation according to the analysis of Edsj{\"o} and Gondolo 
\cite{coann}. %
Present observations favor $h = 0.6\pm 0.1$, and a total matter 
density $\Omega_{M} = 0.3 \pm 0.1$, of which baryons may contribute 
0.02 to 0.08 \cite{cosmparams}. However, we only required that 
$\Omega_{\chi} h^{2} \le 0.5$. We were also interested in models 
where neutralinos
are not the only component of dark matter. In models with 
$\Omega_{\chi} h^{2}\le 0.025$, we rescaled the relevant halo 
densities by a factor
$\Omega_{\chi} h^{2} / 0.025$ to account for the fact that a 
supplemental source of
dark matter is required in such models.

\section{Flux of Gamma--rays from Neutralino Annihilation} 
\label{sec:sensitivity}

\begin{figure*}[t]
\centerline{
\epsfig{file=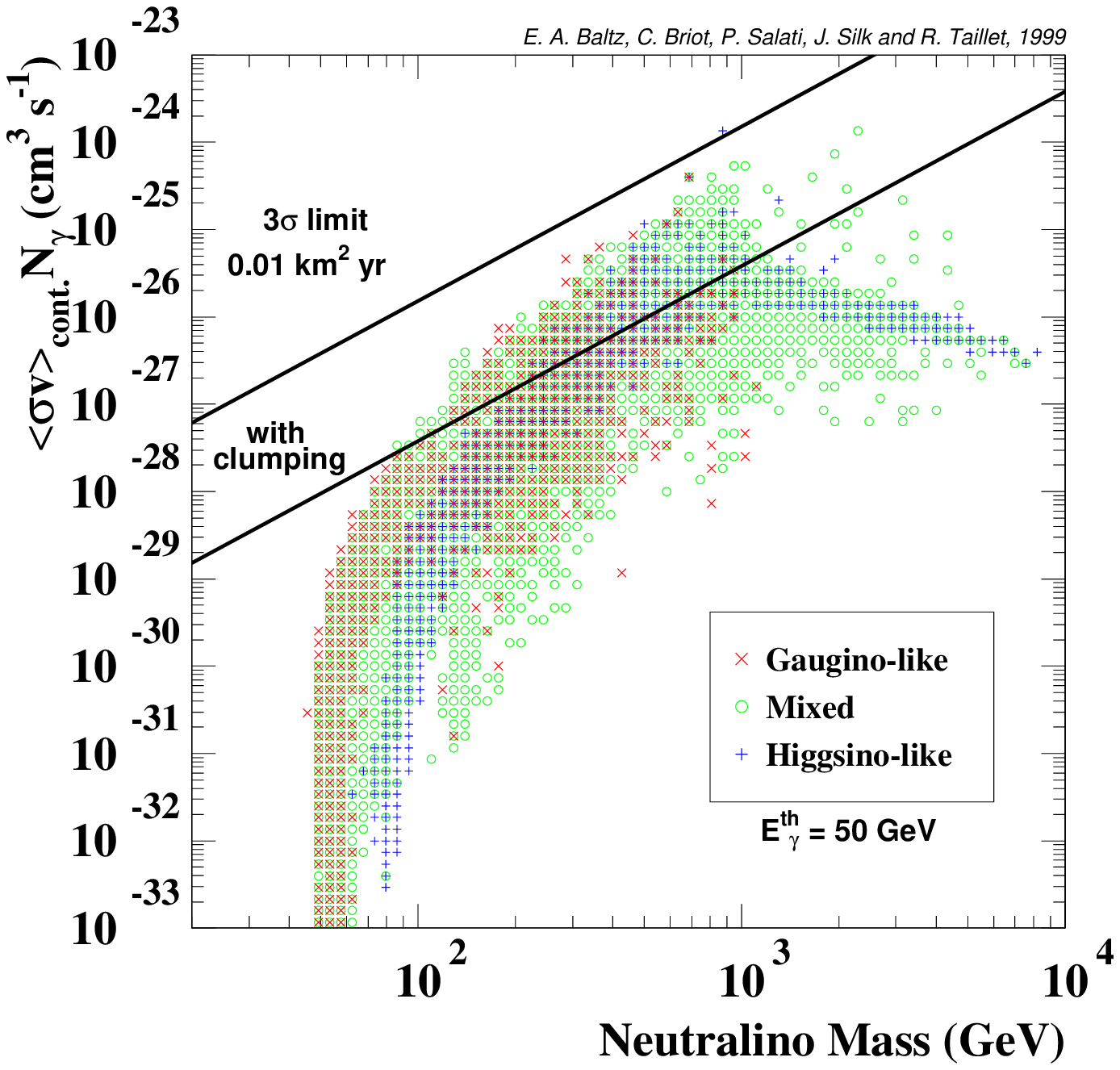,width=0.5\textwidth} 
\epsfig{file=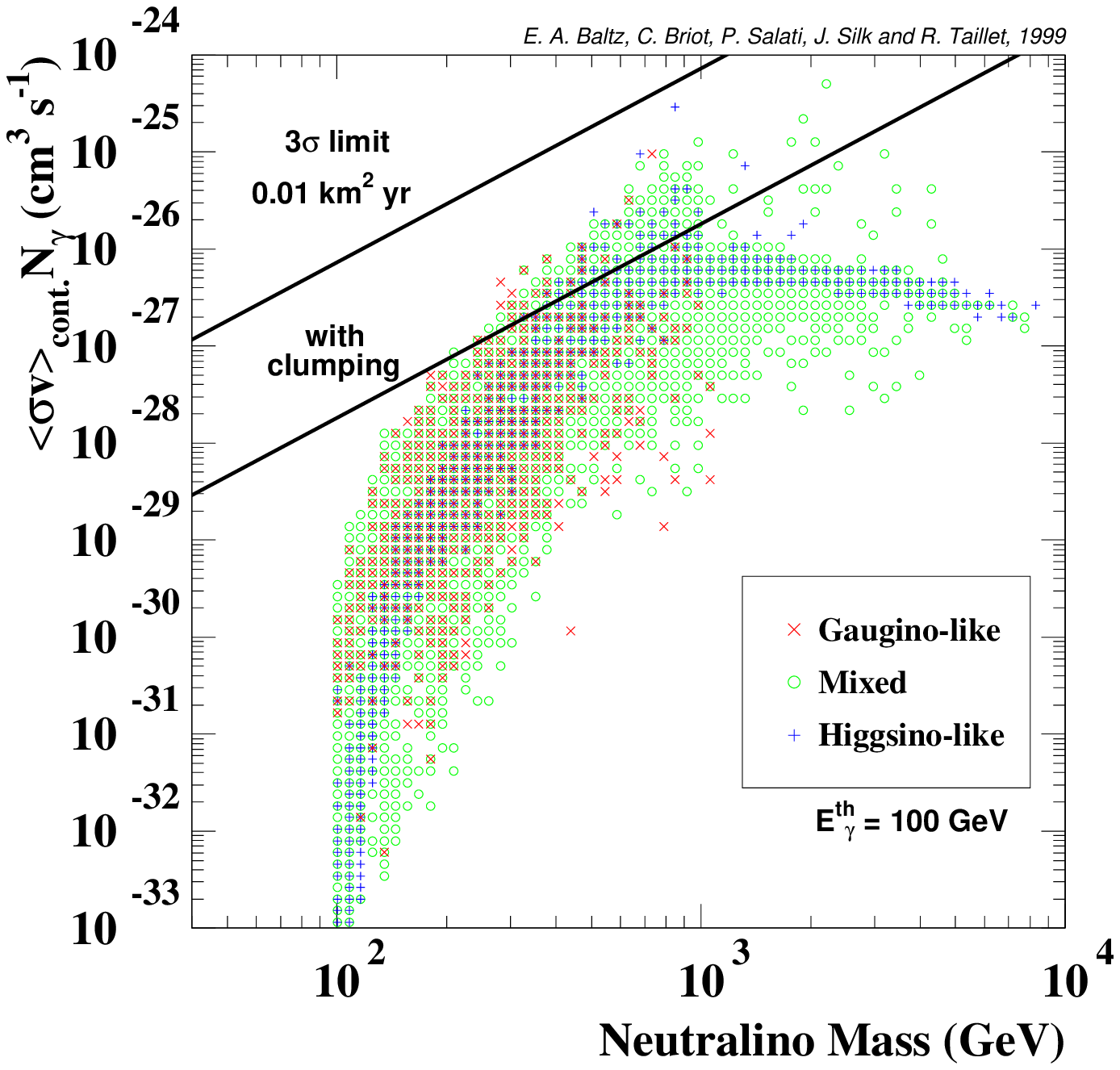,width=0.5\textwidth}} 

\centerline{\epsfig{file=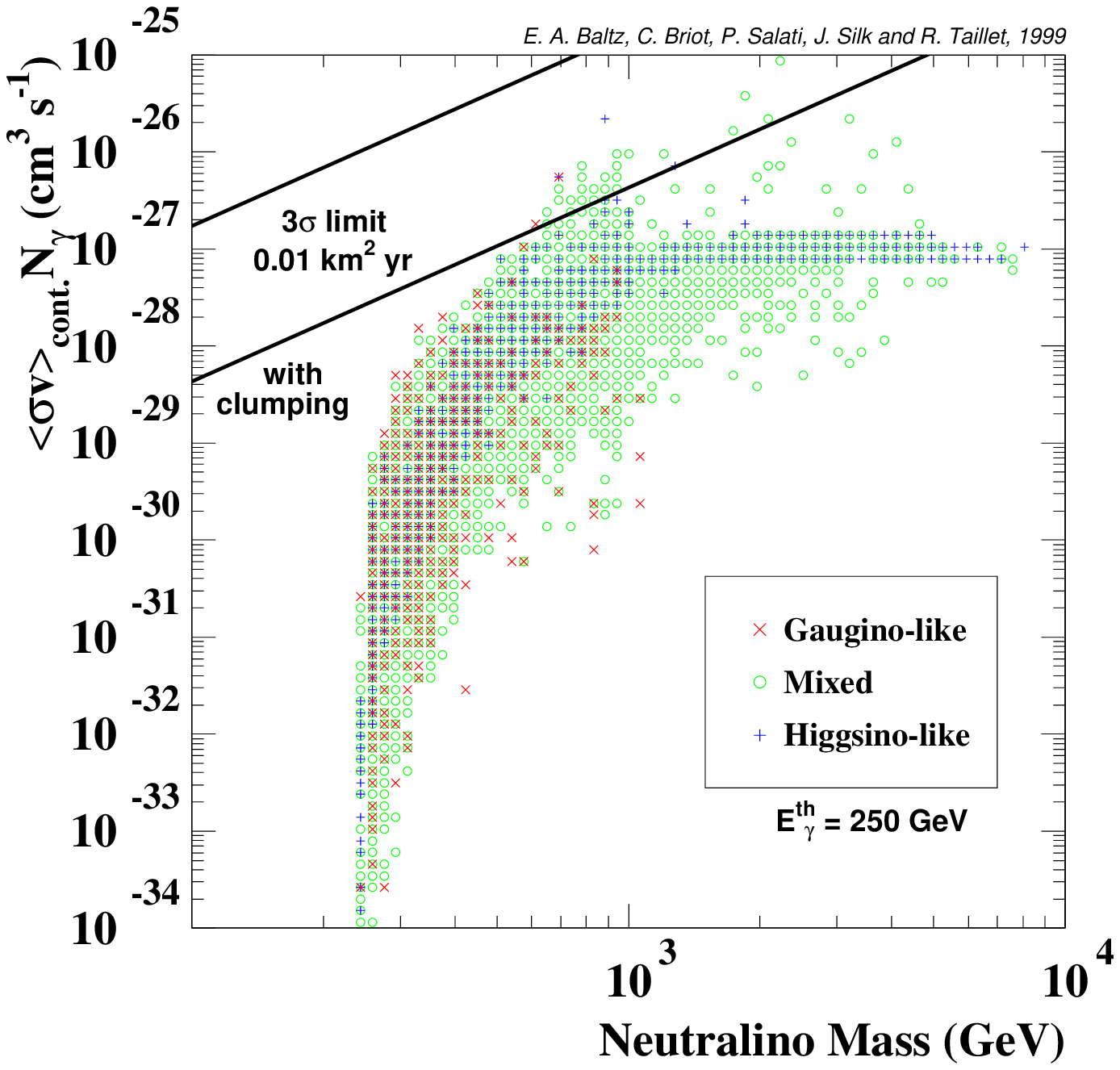,width=0.5\textwidth}} 
\caption{
Annihilation rates in the continuum channels. Three thresholds are 
illustrated, $E_{\rm th}$ = 50, 100 and 250 GeV. Considering M87 as 
the source, the 3--$\sigma$ detection limits for exposures of 0.01 
km$^{2}$ yr are also presented. The regions below the higher heavy 
solid lines will not be accessible, even with the next generation of 
\v{C}erenkov telescopes. The lower solid lines show the region of 
accessibility if annihilation rate is enhanced by a factor of 40 due 
to clumpiness. }
\label{fig:cont}
\end{figure*}

We now present the results of the scans over the supersymmetric 
parameter space.
We first made scatter plots of the continuum gamma--ray production 
rate $\langle \sigma v \rangle_{\rm cont.} \, N_{\gamma}$ versus the 
neutralino mass $m_{\chi}$, for three gamma--ray thresholds $E_{\rm 
th}$ = 50, 100 and 250 GeV. The results are presented in 
Fig.~\ref{fig:cont}, along with the 3--$\sigma$
detection limit of the signal from M87, for a typical exposure of 
$0.01$ km$^2$ yr.
This acceptance corresponds to the next generation of ACTs. The HESS 
project \cite{hess} for instance should reach a threshold of 40 GeV 
for a collecting area of 300 m by 300 m. Its angular resolution 
should be 0.1$^{\circ}$. We have assumed a generous 0.1 yr 
integration time which would correspond to a few months of continuous 
observation. This compares to the VERITAS project which should be 
sensitive to the energy range extending from 50 GeV up to 50 TeV. A 
collecting area of 10,000 m$^{2}$ is expected at 100 GeV, increasing 
by an
order of magnitude for TeV photons. VERITAS should reach an angular 
resolution of 5 arcmin at 100 GeV \cite{veritas}. As shown in the 
left panel of Fig.~\ref{fig:signal_to_noise}, the optimal beam size 
is 1.4 arcmin in the direction
of M87. For a 6 arcmin angular resolution and a 50 GeV threshold, the 
significance
drops down to $S = 0.2$. A 3--$\sigma$ detection level translates 
therefore into the gamma--ray production cross section $\langle 
\sigma v \rangle_{\rm cont.} \, N_\gamma \, = \, 1.5 \times 10^{-24}$ 
cm$^{3}$ s$^{-1}$ for a 1 TeV neutralino. The heavy solid line in the 
upper--left panel of Fig.~\ref{fig:cont} corresponds to a sensitivity 
level of
\beq
\langle \sigma v \rangle_{\rm cont.} \, N_\gamma \geq 1.5 \times 
10^{-26} \, {\rm cm}^{3} \, {\rm s}^{-1} \; \left( 
\frac{m_{\chi}}{100 \; {\rm GeV}} \right)^{2} \;\; . \eeq
The region below that line will not be accessible, even with the next 
generation
of \v{C}erenkov telescopes. When the threshold increases to 100 and 
250 GeV, the sensitivity respectively reaches down a level of $7.2 
\times 10^{-27}$ and $2.8 \times 10^{-27}$ cm$^{3}$ s$^{-1}$. Notice 
that even for a 250 GeV threshold, the number of background photons 
collected within 6 arcmin from the center of M87 amounts to $\sim$ 
4,600 particles. A 3--$\sigma$ signal corresponds to $\sim$ 200 
additional gamma--rays from that hot spot. For a 50 GeV threshold, 
$\sqrt{N\un{bg}} \sim 1.4 \times 10^{5}$ photons whereas $N_{\rm s} 
\sim 1,100$ photons.

\begin{figure*}[t]
\centerline{\epsfig{file=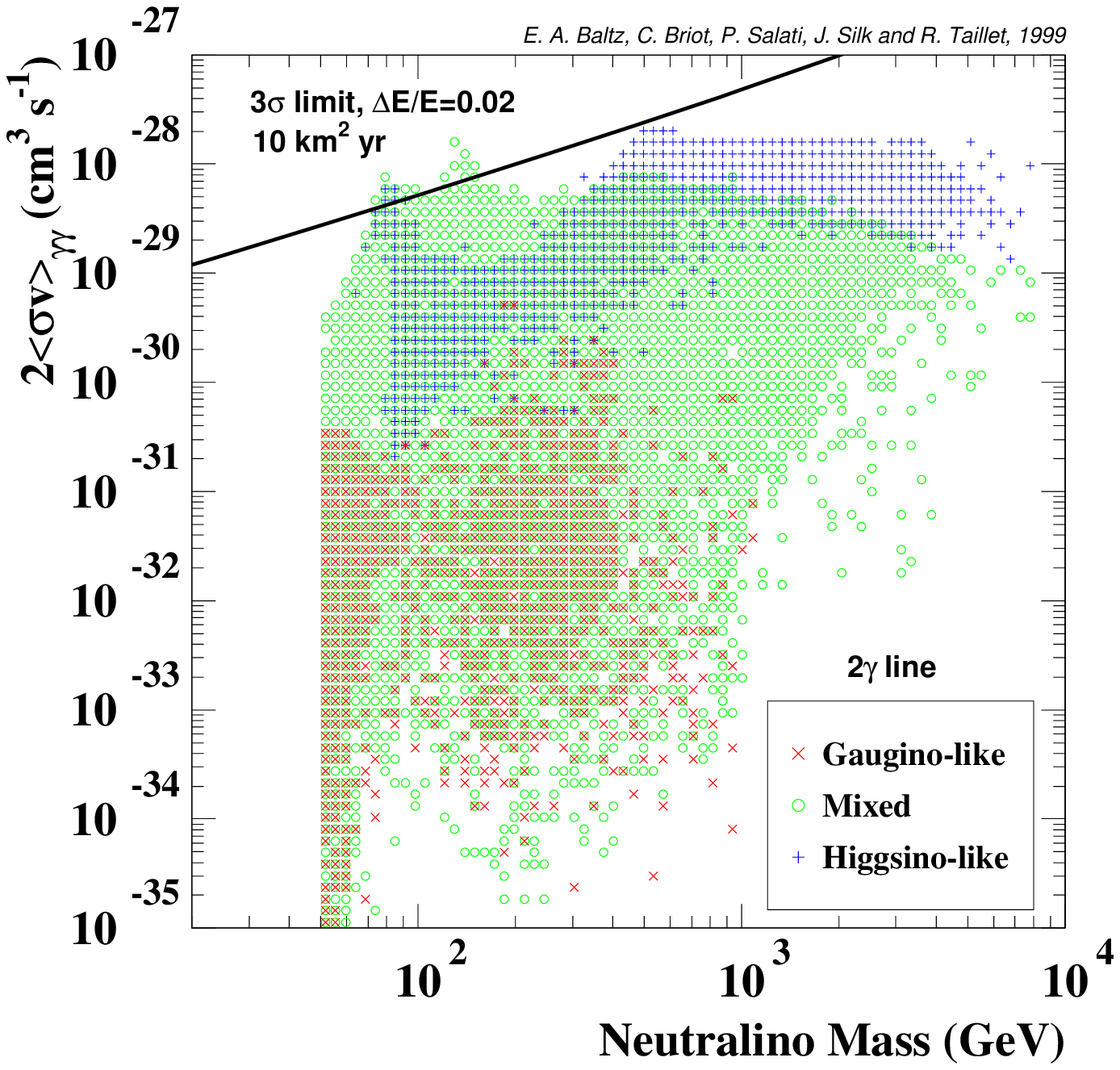,width=0.5\textwidth} 
\epsfig{file=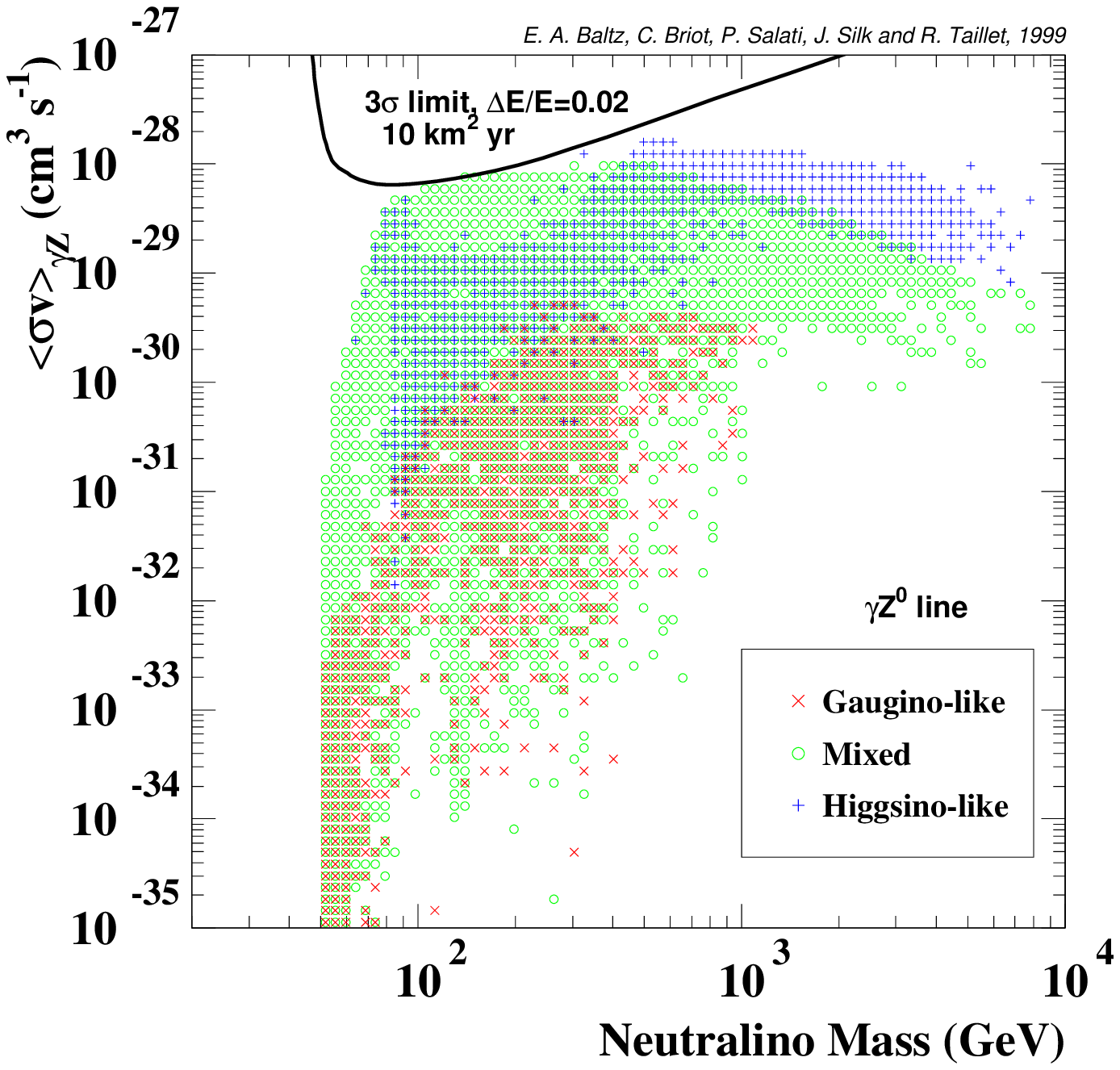,width=0.5\textwidth}} \caption{
Annihilation rates in the monochromatic channels. ({\em left}): 
2--$\gamma$ line.
({\em right}): $Z^{0}\gamma$ line.
Both processes are present with their 3--$\sigma$ detection limits 
for a 10 km$^{2}$ yr exposure towards M87 together with an energy 
resolution of $\Delta E / E = 0.02$.
}
\label{fig:lines}
\end{figure*}

\vskip 0.2cm
In Fig.~\ref{fig:lines}, we present the results for the gamma--ray 
lines. We considered the two processes
$\chi \chi \rightarrow \gamma \gamma$ and $\chi \chi \rightarrow 
Z^{0} \gamma$.
The photon energy in the $2 \gamma$ process is clearly $E_{\gamma} = 
m_{\chi}$, while in the $Z^{0} \gamma$ reaction the photon energy is
\beq
E_{\gamma} \; = \; m_{\chi} \, - \,
{\displaystyle \frac{m_{Z}^{2}}{4 m_{\chi}}} \;\; . \eeq
As in Fig.~\ref{fig:cont}, we also present detection limits. The left 
panel features the two gamma--ray line whereas the right panel 
presents the $Z^{0} \gamma$ process. The sensitivity limits are 
identical for both lines at high energy, \ie, for massive 
neutralinos. In the case of the $Z^{0} \gamma$ line, the photon 
energy becomes vanishingly small as the neutralino mass $m_{\chi}$ 
tends to $m_{Z}/2$. Because the signal becomes swamped inside a very 
strong low--energy gamma--ray background, the sensitivity drops 
completely and the detection limit of the right panel exhibits a 
sharp increase around $m_{Z}/2$. Below that threshold, the $Z^{0} 
\gamma$ process is no longer kinematically allowed. There is an 
additional restriction arising from the energy threshold $E_{\rm th}$ 
of the \v{C}erenkov telescope itself. Requiring that $E_{\gamma} > 
E_{\rm th}$ implies that the neutralino mass should exceed
\beq
m_{\chi} \geq
{\displaystyle
\frac{E_{\rm th} \, + \, \sqrt{m_{Z}^{2} + E_{\rm th}^{2}}}{2}} \;\; 
. \eeq
For a 50 GeV threshold, this translates into $m_{\chi} \geq$ 77 GeV. 
We find that the lines are much more difficult to detect. The 
sensitivities presented here correspond to an exposure of 10 km$^{2}$ 
yr and an energy resolution $\Delta E / E$ = 0.02, both of which are 
unreasonable with today's detectors.

\section{Discussion and Conclusions}
\label{sec:conclusion}

As is clear from Fig.~\ref{fig:cont} and \ref{fig:lines}, the 
supersymmetric parameter space will mostly remain below the 
sensitivity level of the next generation of instruments. A few 
configurations are potentially detectable provided that the ACT 
threshold is decreased down to 50 GeV. %
If neutralinos are clumped inside the dark matter halo of M87, the 
situation considerably improves. As the gamma--ray production rate 
goes as the square of the density, neutralinos annihilate more 
efficiently as they condense. That overall effect may be described by 
the clumpiness factor ${\cal C}$, which is defined as the increase of 
the global annihilation rate that results from a possible clumpy 
structure of the dark matter distribution.

\vskip 0.2cm
Neutralinos are cold dark matter species. As such, they exhibit 
density fluctuations
\beq
\left( {\displaystyle \frac{\delta \rho}{\rho}} \right)^{2} \sim 
\left| \delta_{k} \right|^{2} \, k^{3} \;\; , \eeq
that are related to the comoving wave vector $k$ through \cite{davis} 
\beq
\left| \delta_{k} \right|^{2} \; \equiv \; P(k) = \frac{A k}{(1 + 
\alpha k + \beta k^{1.5} + \gamma k^2 )^2} \;\; . \eeq
with $\alpha = 1.71 \times l$, $\beta = 9 \times l^{1.5}$ and $\gamma 
= l^2$ where $l = (\Omega h^2)^{-1}$. Normalization to $\sigma_8 = 
0.8$ gives $A= 2.82 \times 10^6 \; \mbox{Mpc}^4$ when $\Omega=1$ and 
$h=0.5$. In a restricted wavelength range, it is approximated by a 
power law \beq
P(k) \propto k^{n} \;\; .
\eeq
The power spectrum $P(k)$ of density fluctuations behaves as $k^{-3}$ 
on small scales, \ie, for structures typically lighter than $10^{8}$ 
$M_{\odot}$.
As regards a possible clumpy structure of the halo around M87, the 
relevant mass range extends from $M_{\rm i} \sim 10^{8}$ $M_{\odot}$ 
up to $M_{\rm s} \sim 10^{13}$ $M_{\odot}$. The corresponding 
spectral index $n$ goes from $-2.6$ to $-2.1$.
As shown below, structures smaller than $M_{\rm i}$ turn out to all
have the same density. They contribute identically to the clumpiness
factor ${\cal C}$. There is no larger structure than the halo itself whose 
mass $M_{\rm s}$ reaches $10^{13}$ $M_{\odot}$ in the inner 100 kpc. 
Because the comoving wave vector $k$ scales as $M^{-1/3}$, neutralino 
density fluctuations depend on both the scale $M$ and the redshift 
$z$ as \beq
{\displaystyle \frac{\delta \rho}{\rho}} \propto \left( 1 + z 
\right)^{-1} \, M^{- (n+3)/6} \;\; . \eeq
The redshift factor $\left( 1 + z \right)^{-1}$ is typical of the 
$t^{2/3}$ growth
of density fluctuations in a flat matter--dominated universe. Notice 
that small scale perturbations, for which $n = -3$, all become 
non--linear at the same time.
Their subsequent collapse leads to virialized structures whose 
densities have been enhanced by a factor of $\sim$ 180 with respect 
the epoch of formation, when $\delta \rho / \rho$ reached unity. 
Small scale dark matter clumps all have therefore the same density 
today. The formation redshift of larger structures behaves as
\beq
\left( 1 + z_{\rm F} \right) \propto M^{- (n+3)/6} \;\; , \eeq
so that today, neutralino clumps with mass above $\sim$ $10^{8}$ 
$M_{\odot}$ have a density
\beq
\rho(M) \propto 180 \, \left( 1 + z_{\rm F} \right)^{3} \propto M^{- 
(n+3)/2} \;\; .
\label{density_clump}
\eeq
The density $\rho (M_{\rm s})$ of the largest possible clump should 
be comparable
to the average dark matter density $\rho_{\rm DM}$ in the halo around 
M87.
The distribution of clumps should follow the Press--Schechter's law 
\beq
\frac{dN}{dM} \; = \; \frac{M_{0}}{M^{2}} \;\; . \label{PS_clump}
\eeq
The normalization mass $M_{0}$ obtains from the requirement that the 
clumps make up a fraction $f$ of the halo.
Disregarding for the moment clumps with mass less than $10^{8}$ 
$M_{\odot}$, we get
\beq
M_{0} \; = \; {\displaystyle \frac{f \, M_{\rm s}} {\ln \left( M_{\rm 
s} / M_{\rm i} \right)}} \;\; . \eeq
Some clumps are actually destroyed through the tidal stripping 
resulting from both their mutual interactions and the action of the 
gravitational field of M87. Just like globular clusters orbiting the 
Milky Way, they evaporate so that a fraction $f$ only of the initial 
population is expected to survive.
Inside a clump with mass $M$, the annihilation rate of neutralinos is 
$\langle \sigma v \rangle \, \left\{ \rho (M) / m_{\chi} 
\right\}^{2}$ per unit volume. A net number
$\langle \sigma v \rangle \, \rho (M) \, M / m_{\chi}^{2}$ of 
annihilations take place in the clump per unit time. We infer that 
the total annihilation rate of clumped neutralinos is obtained from 
the convolution \beq
\Gamma_{\rm clump} \; = \;
{\displaystyle \int_{\displaystyle M_{\rm i}}^{\displaystyle M_{\rm 
s}}} \, {\displaystyle \frac{\langle \sigma v \rangle}{m_{\chi}^{2}}} 
\, \rho (M) \, M \;
{\displaystyle \frac{dN}{dM}} \; dM \;\; . \eeq
Taking into account the mass--density relation~(\ref{density_clump}) 
as well as the mass distribution (\ref{PS_clump}) of the clumps, we 
readily infer the rate
\beq
\Gamma_{\rm clump} \; = \;
{\displaystyle \frac{\langle \sigma v \rangle}{m_{\chi}^{2}}} \, 
\rho_{\rm DM} \, M_{0} \,
{\displaystyle \int_{\displaystyle M_{\rm i}}^{\displaystyle M_{\rm 
s}}} \, \left( {\displaystyle \frac{M}{M_{\rm s}}} \right)^{- 
(n+3)/2} \, {\displaystyle \frac{dM}{M}} \;\; .
\eeq
This may be compared to what a homogeneous distribution would yield
\beq
\Gamma_{\rm hom} \; = \;
{\displaystyle \frac{\langle \sigma v \rangle}{m_{\chi}^{2}}} \, 
\rho_{\rm DM} \, M_{\rm s} \;\; .
\eeq
The clumpiness factor ${\cal C}$ may be understood as the enhancement 
ratio $\Gamma_{\rm clump} / \Gamma_{\rm hom}$. Clumps span less space 
than if their matter was homogeneously distributed. In their 
interiors, neutralinos nevertheless annihilate much more efficiently. 
The net effect is the increase
\beq
{\cal C} \; = \;
\left( {\displaystyle \frac{2}{n+3}} \right) \, \left( 
{\displaystyle \frac{f}{\ln \left( M_{\rm s} / M_{\rm i} \right)}} 
\right) \,
\left[ \left( \frac{M_{\rm s}}{M_{\rm i}} \right)^{(n+3)/2} \, - \, 
1 \right] \;\; .
\eeq
This expression gives ${\cal C} \approx 34 \, f$ for $n=-2.1$ and 
${\cal C} \approx 4 \, f$ for $n=-2.6$.
With a general power spectrum $P(k)$, this expression reads 
\begin{equation}
{\cal C} \; = \;
\left( {\displaystyle \frac{f}{\ln \left( M_{\rm s} / M_{\rm i} 
\right)}} \right) \,
\int _{M_i}^{M_s} \left( \frac{M}{M_s} \right)^{-3/2} \left(
\frac{P(M)}{P(M_s)} \right)^{3/2} \frac{dM}{M} \end{equation}
where we go from $P(k)$ to $P(M)$ through \begin{equation}
k = \left( \frac{M}{\msol} \frac{G}{3 H_0^2 \pi^2} \right)^{-1/3} 
\end{equation}
The {\sc CDM} power spectrum leads to ${\cal C} \approx 13 \, f$.
If we now assume that most of the clumps are small and that their 
mass does not exceed $M_{\rm i} = 10^{8}$ $M_{\odot}$, we find \beq
\Gamma_{\rm clump} \; = \;
{\displaystyle \frac{\langle \sigma v \rangle}{m_{\chi}^{2}}} \, \rho 
(M_{\rm i}) \, {\displaystyle \int} \, M \, dN \;\; . \eeq
If that population of light clumps accounts for a fraction $f$ of the 
dark matter halo around M87, the previous relation translates into
\beq
\Gamma_{\rm clump} \; = \;
{\displaystyle \frac{\langle \sigma v \rangle}{m_{\chi}^{2}}} \, \rho 
(M_{\rm i}) \, f \, M_{\rm s} \;\; . \eeq
The clumpiness factor becomes
\beq
{\cal C} \; = \; f \, \frac{\rho(M_i)}{\rho(M_s)} \; = \; f \, \left( 
\frac{M_{\rm s}}{M_{\rm i}} \right)^{(n+3)/2} \;\; , \eeq
in the case of a power-law spectrum or more generally \beq
{\cal C} \; = \; f \,
\left( \frac{M_{\rm s}}{M_{\rm i}} \right)^{3/2} \left(
\frac{P(M_i)}{P(M_s)} \right)^{3/2} \;\; , \eeq
It reaches a value of ${\cal C} \sim 40 \, f$ for the {\sc CDM} power 
spectrum. Varying the fraction $f$ between 0.1 and 1, we conclude 
that depending on the typical size of the clumps, the gamma--ray 
production rate may be enhanced by factors as large as 40. The lower 
solid lines in Fig.~\ref{fig:cont} show the sensitivity limits of 
ACTs assuming that ${\cal C}=40$. A \v{C}erenkov telescope operating 
with a 50 GeV threshold would detect a neutralino--induced gamma--ray 
emission from the giant elliptical galaxy M87 for a part of the 
supersymmetric configurations outlined in the upper--left panel of 
Fig.~\ref{fig:cont}. Even with the annihilation rate enhanced by a 
factor of 40, the gamma ray lines are out of reach. 

\acknowledgments
We wish to thank E.~Nuss for the information which he provided to us 
as well as for stimulating discussions. We thank D. Finkbeiner for 
assistance in using the dust maps of \cite{dustmap}. During his 
visits to Annecy, E.~Baltz has been supported in part by the 
Programme National de Cosmologie. At Berkeley E.~Baltz is supported 
by grants from NASA and DOE.


\end{document}